\documentclass[prl,aps,twocolumn,showpacs,floatfix]{revtex4}
\usepackage{amsmath}
\usepackage{graphicx}

\def\fig_width{3. in} % width of single column figure in PR

\def\prn#1{{\left(#1\right)}}
\def\brk#1{{\left[#1\right]}}

\def\abs#1{{\left|#1\right|}}

\begin{document}

\title{Magnetoelectric Jones Dichroism in Atoms}

\author{D.\ Budker}

\author{J.\ E.\ Stalnaker}

\affiliation{Department of Physics, University of California at
Berkeley, Berkeley, California 94720-7300}

\affiliation{Nuclear Science Division, Lawrence Berkeley National
Laboratory, Berkeley, California 94720}

\date{\today}

\begin{abstract}
The authors suggest that atomic experiments measuring the
interference between magnetic-dipole and electric-field-induced
electric-dipole transition amplitudes provide a valuable system
to study magnetoelectric Jones effects.
\end{abstract}

\pacs{42.25.Lc,32.60.+i}

\maketitle

A recent letter \cite{roth00} reported the first observation of
magnetoelectric Jones birefringence in liquids (see also Ref.\
\cite{rikken01}).  This observation helped to clarify some of the
long-standing theoretical confusion surrounding Jones
birefringence and the associated Jones dichroism (collectively
known as Jones effects) \cite{graham83,ross89}. The interest in
further understanding these effects has led to the investigation
of other experimental systems which may exhibit Jones effects.
These include the possibility of observing the effects through
beam divergence in uniaxial crystals \cite{izdebski01} and
possible observation in the quantum vacuum \cite{rikken00}.  In
this letter we point out that Jones dichroism can be studied in
atomic systems under much less severe experimental requirements.
In addition, these atomic systems are more amenable to theoretical
analysis than the relatively complicated condensed-matter systems
that have been studied to date.  The simplicity of these systems
may help to expand the understanding of the manifestation of
Jones effects in general. We also point out that our recent
experiment \cite{stalnaker02} measuring interference between
magnetic-dipole and electric-field-induced electric-dipole
transition amplitudes in atomic ytterbium constitutes a
measurement of Jones dichroism in a simple atomic system.

The development of the Jones matrix calculus for describing the
propagation of light led to the prediction of two distinct types
of linear birefringence and dichroism \cite{jones48}.  The two
types of effects differ in the orientation of the birefringent
and dichroic axes.  The Jones formalism revealed that certain
uniaxial media may exhibit birefringence and dichroism along axes
which are at $\pm 45^\circ$ relative to the axis of anisotropy.
Birefringent and dichroic effects of this type are called Jones
effects. They are distinct from the familiar birefringence and
dichroism, which have axes parallel and perpendicular to the axis
of anisotropy.

There has been theoretical discussion concerning the requirements
for media that may exhibit Jones effects and what transition
moments must be accounted for in order to describe it
\cite{graham83,ross89}.  In Ref.\ \cite{graham83}, it was shown
that Jones effects may be induced in isotropic media by the
application of parallel electric and magnetic fields.  If the
direction of light propagation is perpendicular to the electric
and magnetic fields, the Jones effects are described by
\begin{equation}
\Delta n_J \equiv n_{+45^\circ}-n_{-45^\circ},
\end{equation}
where $n_{\pm 45^\circ}$ is the complex index of refraction for
light polarized at $\pm 45^\circ$ relative to the electric and
magnetic fields.  The real and imaginary parts of $\Delta n_J$
describe Jones birefringence and dichroism, respectively.  On a
microscopic level, Jones dichroism may manifest itself as a
difference between the rates with which atoms of the medium are
transferred to the excited state in the presence of light
polarized at $\pm45^\circ$ relative to the electric and magnetic
fields given by
\begin{equation}
\Delta \Gamma_J \equiv \Gamma_{+45^\circ}-\Gamma_{-45^\circ}.
\end{equation}

Jones effects generally occur in materials which exhibit the more
familiar birefringence and dichroism.  In addition, Jones effects
are predicted to be significantly smaller than the usual
birefringence and dichroism in most media.  Consequently,
magnetoelectric Jones birefringence has been observed only
recently in molecular liquids under extreme experimental
conditions \cite{roth00,rikken01}.

To our knowledge, the observation of Jones dichroism has not been
reported as such.  Here we point out that Stark-interference
experiments \cite{bouchiat74} which utilize parallel electric and
magnetic fields provide a simple atomic system which exhibits
Jones dichroism.  The experiment \cite{stalnaker02} measuring a
highly forbidden magnetic-dipole transition amplitude in atomic
ytterbium using this technique constitutes such a system and its
results can be interpreted as an observation of magnetoelectric
Jones dichroism.

In the experiment \cite{stalnaker02}, we studied a highly
forbidden transition between states of the same parity.  In the
absence of external fields and neglecting parity-nonconserving
effects, the transition occurs only through a small
magnetic-dipole amplitude ($\approx 10 ^{-4} \ \mu_B$, where
$\mu_B$ is the Bohr magneton). By applying a static electric
field, an electric-dipole amplitude is induced through mixing of
opposite-parity states. An atomic beam of ytterbium was excited
with resonant laser light propagating perpendicularly to parallel
electric and magnetic fields.  The excitation light was polarized
at an angle $\theta$ relative to the external fields. For the
transition studied in our experiment (between a ground state with
total angular momentum equal to zero and an excited state with
total angular momentum equal to one), the electric field,
${\mathbf E}$, results in a Stark-induced electric-dipole
transition amplitude to the $M_J'$ magnetic sublevel of the
excited state given by \cite{bouchiat74}
\begin{equation}
 A\prn{E1_{St}} = i\, \beta \left( {\mathbf E} \times {\boldsymbol \varepsilon} \right)_{-M'_{J}}, \label{StarkAmp}
\end{equation}
where ${\boldsymbol \varepsilon}$ is the electric-field amplitude
of the laser light, $({\mathbf E} \times
{\boldsymbol\varepsilon})_{-M'_{J}}$ is the $-M'_J$ component of
the vector in the spherical basis, and $\beta$ is the vector
transition polarizability \cite{note}. The magnetic-dipole
transition amplitude is given by
\begin{equation}
A(M1)= \mu ({\bf \hat k} \times {\boldsymbol
\varepsilon})_{-M_J'},
\end{equation}
where ${\bf \hat k}$ is the direction of propagation of the
excitation light, ${\bf \hat k} \times {\boldsymbol \varepsilon}$
is the magnetic-field amplitude of the light, and $\mu$ is the
magnetic-dipole matrix element between the ground state and any
of the $M_J'$ magnetic sublevels of the excited state. The
transition rate is therefore
\begin{equation}
\begin{split}
\Gamma & \propto \sum_{M_J'} \abs{A\prn{E1_{St}} + A\prn{M1}}^2\\
& \propto \sum_{M_J'} \abs{A\prn{E1_{St}}}^2 + 2 {\rm
Re}\brk{A\prn{E1_{St}}
  {A\prn{M1}^{\ast}}} + \abs{A\prn{M1}}^2. \label{rate}
\end{split}
\end{equation}
As is discussed in Ref. \cite{stalnaker02}, the interference term
in Eq. \eqref{rate} is of opposite sign for the $M_J'=+1$ and
$M_J'=-1$ magnetic sublevels.  It is therefore necessary to apply
a magnetic field to resolve the different sublevels in order to
observe the effect of this term. The signal due to the
interference term is proportional to the rotational invariant
\begin{equation}
[( {\bf E} \times {\boldsymbol\varepsilon} ) \times ( {\bf\hat k}
\times {\boldsymbol\varepsilon} )] \cdot {\bf\hat B},
\label{interference}
\end{equation}
which is also true in a more general case where ${\bf E}$ and
${\bf B}$ are not necessarily collinear.

\begin{figure}
\includegraphics[width=2.75 in]{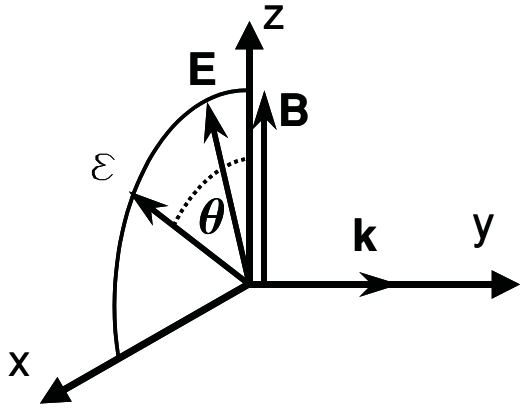}
\caption{Orientation of external fields.} \label{geometry}
\end{figure}

We define the $z$ axis to be along the direction of the magnetic
field, ${\bf B} = B \, {\bf \hat z}$, and define the $x$ axis so
that the electric field lies in the $x$-$z$ plane, ${\bf E}=E_x
\, {\bf\hat x}+E_z \, {\bf\hat z}$. We assume that the light
propagation is perpendicular to both fields, ${\bf k} = k \,
{\bf\hat y}$ and that the light is linearly polarized at an angle
$\theta$ relative to the magnetic field, ${\hat \varepsilon} =
{\rm sin}\theta \, {\bf\hat x} + {\rm cos}\theta \, {\bf\hat z}$
(Fig. \ref{geometry}).
The $\abs{A\prn{E1_{St}}}^2$ and $\abs{A\prn{M1}}^2$ terms in Eq.
\eqref{rate} are independent of the sign of the angle of
polarization while the interference term is odd with $\theta$.
Using expression \eqref{interference} it is easily shown that the
difference in transition rates for $\pm \theta$ results in a
Jones dichroism given by
\begin{equation}
\Gamma_{+\theta}-\Gamma_{-\theta} \propto E_z \; {\rm sin}\theta
\; {\rm cos}\theta. \label{gammatheta}
\end{equation}
The factor $E_z$ in Eqn. \eqref{gammatheta} shows that
\begin{equation}
\Delta \Gamma_J \propto {\bf E} \cdot {\bf \hat B},
\end{equation}
which is the predicted dependence of the magnetoelectric Jones
effects on ${\bf E}$ and ${\bf B}$ \cite{graham83}.  It is
interesting to note that the transition rate depends on the
magnitude of the magnetic field only for values of the magnetic
field which do not fully resolve the magnetic sublevels.  This is
analogous to the change in magnetic-field dependence of the
resonant Faraday rotation (see for example Ref.\ \cite{budker02}).

\begin{figure}
\includegraphics[width=3.375 in]{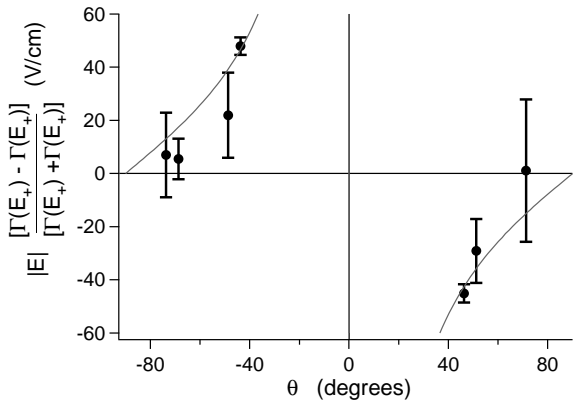}
\caption{Experimental results showing the dependence of the
fractional transition-rate asymmetry, normalized by the magnitude
of the electric field, on the angle of light polarization. Solid
line shows the expected dependence. Data was taken in the work of
Ref. \cite{stalnaker02} and experimental details are contained
therein. } \label{polDep}
\end{figure}

Due to the weakness of the forbidden transition studied, we
determined the transition rate by observing fluorescence in a
decay branch of the excited state rather than detecting
absorption. In our experiment the observed Jones dichroism is
significantly smaller ($\approx 5 \times 10^{-3}$ at the electric
fields used in the experiment) than the normal dichroism, which
is dominated by the Stark-induced amplitude. As can be seen from
Eq.(\ref{StarkAmp}), only the component of the light electric
field that is perpendicular to ${\bf E}$ contributes to the
transition rate.  Thus, the dominant fluorescence signal is
proportional to ${\rm sin}^2\theta$ and to $\abs{E}^2$. These
dependences were verified experimentally. The interference term
responsible for Jones dichroism was isolated from the dominant
signal by comparing the fluorescence spectra for opposite
electric fields.  In the data analysis we normalized the
interference term to the dominant signal resulting in an
asymmetry given by
\begin{equation}
\frac {\Gamma(E_+)-\Gamma(E_-)}{\Gamma(E_+)+\Gamma(E_-)}
 = \frac {2 M1} {\beta E} \frac{{\rm cos}(\theta)} {{\rm
sin}(\theta)} M_J.\label{asymEqn}
\end{equation}
The dependence of the asymmetry on the electric field, magnetic
field, and polarization angle was verified experimentally (see
Ref. \cite{stalnaker02} for figure showing the interference term
versus the magnitude of the electric field). Figure \ref{polDep}
shows the experimental fractional asymmetry [Eq.\eqref{asymEqn}]
plotted versus the polarization angle.  We have normalized the
signal to the magnitude of the electric field in order to combine
data taken a different electric fields and leave only the
polarization angle dependence.  Also shown is the expected angular
dependence of the asymmetry.  Most of the data was taken with
light polarized at $\theta=\pm 45^\circ$ relative to the electric
field since the interference term is maximal at these values [see
Eq. \eqref{gammatheta}].  The difference in the sign of the
asymmetry for $\theta=\pm 45^\circ$ clearly indicates a nonzero
value of $\Delta \Gamma_J$, verifying the key signature of Jones
dichroism.

We note that although the Jones dichroism was significantly
smaller than the usual dichroism in our experiment, it is possible
to significantly increase its size by using an allowed
magnetic-dipole transition.  In fact, it possible to have both
the Jones dichroism and the regular dichroism of the same order
as the overall absorption, which can be substantial in the case
of an allowed magnetic-dipole transition.

Finally, we point out that atomic systems may be of use in
measuring other types of magnetoelectric effects which are
currently being studied in more complicated systems, such as more
common forms of magnetoelectric linear birefringence
\cite{roth02} and magnetoelectric directional anisotropy
\cite{rikken02}.  In fact, expression (\ref{interference}) shows
that this system exhibits both of these effects. It is
interesting to note that a polarization-dependent directional
anisotropy is present for both parallel and perpendicular
electric and magnetic fields.  For the case of perpendicular
electric and magnetic fields a component of the directional
anisotropy is present even when averaged over the polarization
angle.

The authors acknowledge their co-authors in Ref.
\cite{stalnaker02}, D.\ P.\ DeMille, S.\ J.\ Freedman, and V.\ V.\
Yashchuk, and helpful discussions with D.\ F.\ Kimball, A.-T.
Nguyen, A.\ O.\ Sushkov, S.\ M.\ Rochester, and E.\ D.\ Commins.
This work has been supported by NSF, ONR, and the Miller Institute
for Basic Research in Science.

%%%%%%%%%%%%%%%%%%%%%%%%%%%%%%%%%%%%%%%%%%%%%%%%%%%%%%%%%%%%%%%%%%%
%  Figures
%%%%%%%%%%%%%%%%%%%%%%%%%%%%%%%%%%%%%%%%%%%%%%%%%%%%%%%%%%%%%%%%%%%

\end{document}